\documentclass{PoS}

\usepackage{graphics,wrapfig,times}
\usepackage{epsf}
\usepackage{amsmath,amssymb,latexsym,float,algorithm,algorithmic}
 
\newcommand{\R}{\text{\fontshape{n}\selectfont I\kern-.42exR}}
\newcommand{\1}{\text{\fontshape{n}\selectfont 1\kern-.56exl}}

\PoS{PoS(LAT2005)101}
\title{Shifted unitary orthogonal methods for the overlap inversion}
\ShortTitle{The shifted unitary orthogonal method}

\author{\speaker{Artan Bori\c{c}i}
\thanks{The authors participation in the conference and part of this work
are funded from the NATO grant EAP.RIG.981410.}\\
Physics Department\\
University of Tirana\\
Blvd. King Zog I\\
Tirana - Albania\\
E-mail: \email{borici@fshn.edu.al}\\
}

\author{Alban Allko\c{c}i\\
Computer Science Section\\
Polytechnic University of Tirana\\
Mother Theresa Square\\
Tirana - Albania\\
E-mail: \email{alban@fie.upt.al}\\
}

\abstract{In this work we compare the convergence of the shifted
unitary orthogonal method (SUOM) and different Krylov subspace solvers
for propagator computations with overlap fermions.
We show that the SUOM algorithm performs similarly to the
shifted unitary minimal residual method (SUMR) with the latter
converging slightly faster. When the geometric optimality is
applied to SUOM we get e new algorithm which is faster than SUMR.
}
\FullConference{XXIIIrd International Symposium on Lattice Field Theory\\
                25-30 July 2005\\
                Trinity College, Dublin, Ireland}

\begin{document}

\section{Introduction}

In this paper we report on the progress that we have made in the
search for an optimal Krylov subspace method for overlap fermions.
In the previous lattice conference we reported preliminary
results on a new method,
the shifted unitary orthogonal method or SUOM
\cite{Borici_lat04}.
The method is in fact the three-term-recurrence specialisation of the
full orthogonalisation method (FOM)
in case of shifted unitary matrices.
Likewise, the shifted unitary minimal residual method (SUMR),
discovered earlier by \cite{JagelsReichel94,Arnold_et_al}
is a short-recurrence specialisation of the
generalised minimal residual method (GMRES).

Our task is to solve the linear system:
\begin{equation}\label{lin_sys}
D x = b, ~~~~~~x,b \in {\mathbb C}^N, ~~~D \in {\mathbb C}^{N\times N}
\end{equation}
where $D$ is the Neuberger's Overlap operator \cite{Ne98}:
\begin{equation}
D = c_1 I + c_2 V
\end{equation}
Here $c_1 = (1 + m)/2, c_2 = (1 - m)/2$ and $m$ are
the bare fermion mass,
$V$ is a unitary matrix given by:
\begin{equation}
V = D_W (D_W^*D_W)^{-\frac{1}{2}}
\end{equation}
where
$D_W$ is Wilson-Dirac lattice operator.

\section{Arnoldi iteration for unitary matrices}

Some time ago Rutishauser \cite{Rutishauser66} observed
that for upper Hessenberg unitary matrices one can write
$H = L U^{-1}$,
where $L$ and $U$ are lower and upper bidiagonal matrices.
Applying this decomposition for the Arnoldi iteration,
\begin{equation}
V Q_k= Q_k H_k + h_{k+1,k}  q_{k+1} e_k^T
\end{equation}
one obtains:
\begin{equation}
V Q_k U_k = Q_k L_k + h_{k+1,k}  q_{k+1} e_k^T.
\end{equation}
Since $U_k,L_k$ are bidiagonal matrices we arrive to the
following three-term recursion:
\begin{equation}
l_{k+1,k}  q_{k+1} = Vq_k - q_k l_{kk} + V q_{k-1} u_{k-1,k}
\end{equation}
This way we obtain the unitary Arnoldi process shown
in Algorithm \ref{uap}.
\begin{algorithm}
\caption{Unitary Arnoldi Process}
\label{uap}
\begin{algorithmic}
\STATE $q_1 = b / ||b||^2$
\FOR{$~k = 1, \ldots n$}
    \STATE $w_k = V q_k$
    \IF{$k=1$}
       \STATE $u_{k-1,k} = 0$
    \ELSE
       \STATE $u_{k-1,k} = - (q_{k-1}^* w_k) / (q_{k-1}^* w_{k-1})$
    \ENDIF
    \STATE $l_{kk} = q_k^* w_k + w_{k-1} ~u_{k-1,k}$
    \STATE $w_{k+1} = w_k - q_k ~l_{kk} + w_{k-1} ~u_{k-1,k}$
    \STATE $l_{k+1,k} = ||w_{k+1}||_2$
    \IF{$l_{k+1,k} = 0$}
      \STATE stop
    \ENDIF
    \STATE $q_{k+1} = w_{k+1} / l_{k+1,k}$
\ENDFOR
\end{algorithmic}
\end{algorithm}
If we denote ${\tilde H}_k = H_k + l_{k+1,k} e_k e_k^T$, it
is easy to show that for the unitary Arnoldi process we get:
\begin{equation}
{\tilde H}_k^* {\tilde H}_k = H_k^* H_k + l_{k+1,k}^2 e_k e_k^T = I_k
\end{equation}

\section{The SUOM algorithm}

Using the unitary Arnoldi process
one can ask an approximate solution of \ref{lin_sys}
as a linear combination of the
Arnoldi vectors $Q_k$. This leads to solving the smaller
linear system:
\begin{equation}
(c_1 I_k + c_2 L_k U_k^{-1}) y_k = e_1
\end{equation}
The matrix in the left hand side is upper Hessenberg. In order
to solve a simpler system we
define $z_k = U_k^{-1} y_k$ and get:
\begin{equation}
(c_1 U_k + c_2 L_k) z_k = e_1
\end{equation}
where now $T_k = c_1 U_k + c_2 L_k$ is a tridiagonal matrix.
Note that the solution to the above system can be updated
recursively and this way one can update the solution of the
original system by short recurrences. Here we omit the details and
refer the interested reader to a forthcoming publication
\cite{suom_paper}. The resulting algorithm is shown in Algorithm
\ref{suom_algor}.
\begin{algorithm}[htp]
\caption{SUOM algorithm}
\label{suom_algor}
\begin{algorithmic}
\STATE $\rho = ||b||_2; ~q_1 = b/\rho; ~w_1 = q_1$
\STATE $l_{11} = q_1^H V q_1$
\STATE $\tilde{q} = Vq_1 - l_{11} q_1$
\STATE $l_{21} = ||\tilde{q}||_2; ~q_2 = \tilde{q} / l_{21}$
\STATE $\tilde{l}_{11} = c_1 + c_2 l_{11}$
\STATE $\alpha_1 = \rho/\tilde{l}_{11}; ~x_1 = \alpha_1 w_1;
~r_1 = b - \alpha_1 D w_1$
\FOR{$~k = 2,3, \ldots$}
    \STATE $u_{k-1k} = - q_{k-1}^H V q_k / q_{k-1}^H V q_{k-1}$
    \STATE $l_{kk} = q_k^H V q_k + u_{k-1k} q_k^H V q_{k-1}$
    \STATE $\tilde{q} = (V - l_{kk}) q_k + u_{k-1k} V q_{k-1}$
    \STATE $l_{k+1k} = ||\tilde{q}||_2$
    \STATE $q_{k+1} = \tilde{q} / l_{k+1k}$
    \STATE $\tilde{l}_{kk} = c_1 + c_2 l_{kk}
- c_1 c_2 l_{kk-1} u_{k-1k} / \tilde{l}_{k-1k-1}$
    \STATE $\alpha_k = - c_2 l_{kk-1} / \tilde{l}_{kk} \alpha_{k-1}$
    \STATE $w_k = q_k + u_{k-1k} q_{k-1}
 - c_1 u_{k-1k} / \tilde{l}_{k-1k-1} w_{k-1}$
    \STATE $x_k = x_{k-1} + \alpha_k w_k$
    \STATE $r_k = r_{k-1} - \alpha_k D w_k$
    \IF{$||r_k||_2 <$ tol $\rho$}
       \STATE stop
    \ENDIF
\ENDFOR
\end{algorithmic}
\end{algorithm}
Note that $D w_k$ multiplication is redundant. Indeed multiplying
by $D$ both sides of the equation:
\begin{equation}
w_k = q_k + u_{k-1k} q_{k-1}
    - c_1 u_{k-1k} / \tilde{l}_{k-1k-1} w_{k-1}
\end{equation}
and saving the $Vq_k$ and $Vq_{k-1}$ vectors one obtains for free
the $Dw_k$ vector.

The SUOM algorithm is algebraically optimal. This means that the
algorithm constructs a new residual vector which is orthogonal to the
Krylov subspace already in place. However, a more satisfying
optimality is the geometric optimality, which is a feature of
algorithms with a new residual being smaller than the previous one.
In this case one seeks the solution such that the residual vector
norm is minimal:
\begin{equation}
x = \text{arg} \min ||b - D x||_2
\end{equation}
Arnoldi process offers an orthogonal basis vectors which can be used
to project the above large least squares problem into a smaller
problem:
\begin{equation}
{\tilde y}_k = \text{arg} \min ||e_1 - {\tilde H}_k {\tilde y}_k||_2
\end{equation}
The algorithm that is derived this way is a minimal residual algorithm
for shifted unitary matrices. We call it SUOM+ since it is a different
algorithm from SUMR. The latter uses Givens rotations to implement the
isometric Arnoldi process
\cite{JagelsReichel94} as opposed to the unitary Arnoldi process
implementation used in our case, Algorithm \ref{uap}. The full details
of the SUOM+ algorithm can be found in \cite{suom_paper}.

\section{Comparison of algorithms}

In Figure 1 we show
the convergence of various algorithms as a function of
Wilson matrix-vector multiplication number on
$8^316$ lattices at various couplings and quark masses.
For the overlap matrix-vector multiplication
we use the double pass Lanczos algorithm (without small eigenspace
projection of $H_W$)

We show the convergence os SUOM, SUMR,
Conjugate Residuals (CR),
Conjugate Gradients on Normal Equations (CGNE)
and CG-CHI.
The latter is the CGNE which solves simultaneously the decoupled
chiral systems appearing in the matrix $D^*D$. We have preliminary
results for the SUOM+ algorithm in one case only.

{\hspace{-1.5cm}
\includegraphics[width=8cm,height=9cm]{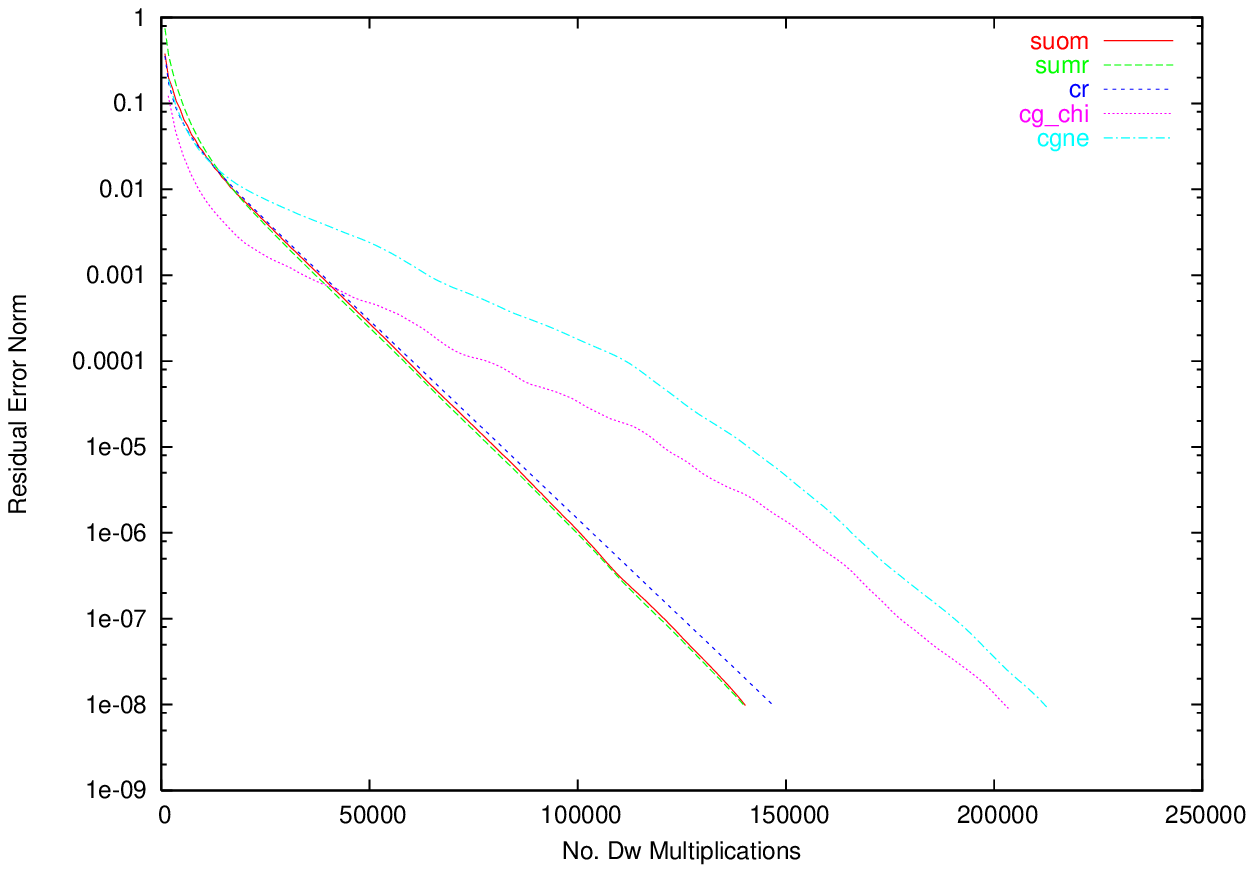}
\includegraphics[width=8cm,height=9cm]{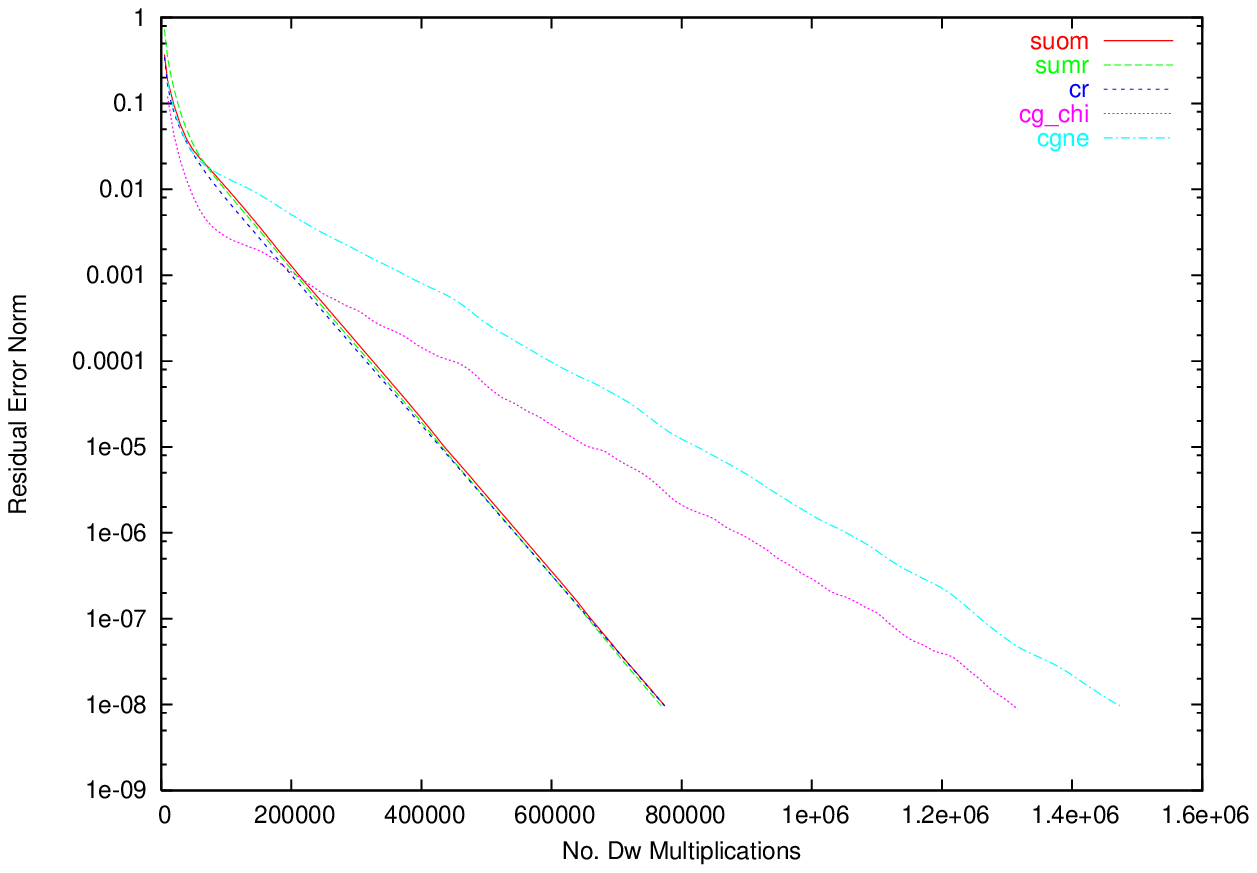}
}

{\hspace{-1.5cm}
\includegraphics[width=8cm,height=9cm]{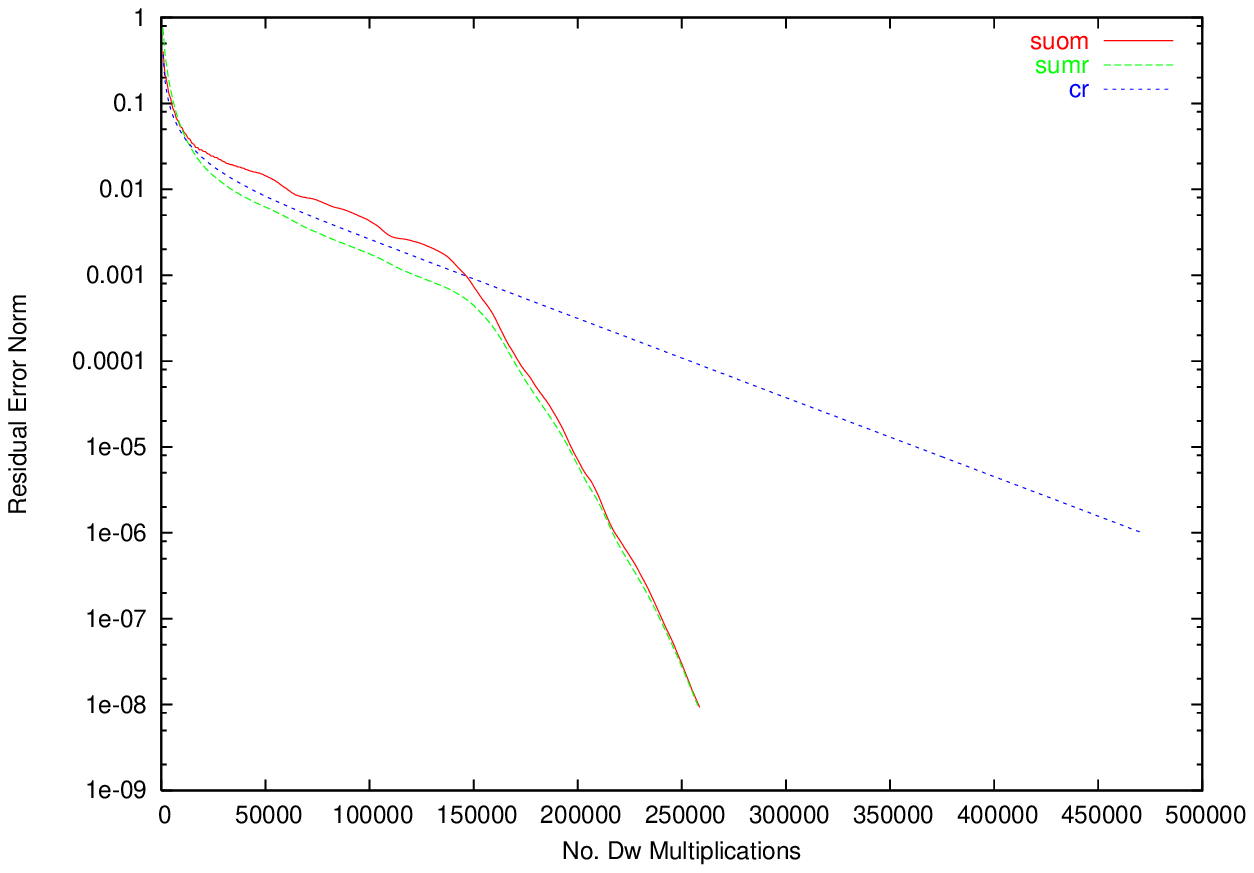}
\includegraphics[width=8cm,height=9cm]{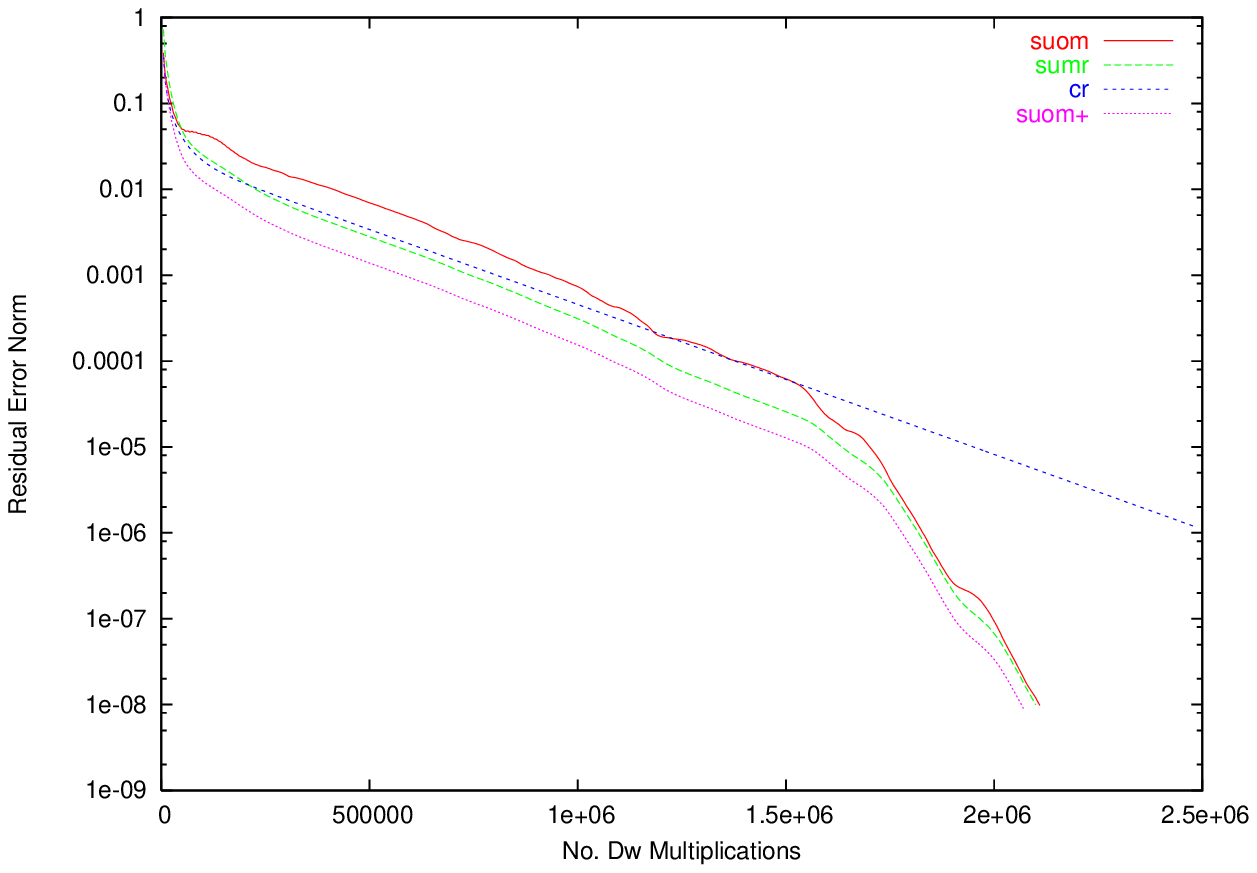}
}
\\Figure 1: Convergence history of various solvers for quark masses
$m=0.05$ (upper panel) and $m=0.01$ (lower panel)
on background gauge fields at $\beta = 6$ (left panel) and
$\beta = 5.7$ (right panel).

\vspace{0.5cm}
The first observation is that for quark propagator computations SUOM,
SUMR and CR are more efficient than CGNE and CG-CHI algorithms.
This is observed by the other groups as well \cite{Arnold_et_al}.
This is why we did not run further these algorithms for smaller quark
masses.

Another interesting observation is that CR converges neck-to-neck with
SUOM and SUMR algorithms for moderate quark masses. As this is lowered,
we see that SUOM and SUMR convergence rate becomes larger than that
of CR at a certain
accuracy threshold, which depends on $\beta$.

Hence, the best algorithms are the optimal algorithms
SUOM and SUMR. These converge in all cases similarly with
SUMR being slightly faster, something which should be expected
from the geometric optimality of SUMR.

However, we see, in the only case available,
that SUOM+ converges faster
than SUOM and SUMR. We expected a different behaviour of SUOM+ from
SUMR, but there are no theoretical grounds to expect that SUOM+
is faster than SUMR.
Since the result is preliminary, further tests are
needed to make a definite conclusion.

\section{Conclusion}

We have shown how to build a class of optimal iterative solvers
for overlap fermions using the unitary Arnoldi process \ref{uap}.
This is easier to implement using the Rutishauser decomposition
\cite{Rutishauser66} than Givens rotations \cite{JagelsReichel94}.

Preliminary results show that SUOM+ algorithm may converge faster
then SUMR. It is expected that different implementations can give
different results and in our case this is obvious. What is not
obviuos is which implementation is faster and this has to be
further investigated \cite{suom_paper}.

\end{document}